\documentclass[prd,amsmath,amssymb,nofootinbib]{revtex4}
\usepackage{color}
\usepackage{ascmac}
\usepackage{feynmp}
\usepackage[pdftex]{graphicx}
\usepackage{graphicx}
\usepackage{ulem}
\usepackage{multirow}
\usepackage{slashbox,pict2e}
\def\x{{\boldsymbol x}}

\def\0{{\boldsymbol 0}}
\def\p{{\boldsymbol p}}
\def\q{{\boldsymbol q}}

\begin{document}

\begin{flushright}
KEK-TH-1776\\
UT-Komaba/14-6
\end{flushright}

\title{Dynamic Critical Exponent from One- and Two-Particle Irreducible $1/N$ Expansions of Effective and Microscopic Theories}
\author{Osamu Morimatsu}
\address{KEK Theory Center, IPNS,
          High Energy Accelerator Research Organization (KEK), \\
           1-1 Oho, Tsukuba, Ibaraki, 305-0801, Japan}
\address{Department of Physics, Faculty of Science, University of Tokyo,\\
7-3-1 Hongo Bunkyo-ku Tokyo 113-0033, Japan}
\address{Department of Particle and Nuclear Studies, \\
Graduate University for  Advanced Studies (SOKENDAI), 
1-1 Oho, Tsukuba, Ibaraki 305-0801, Japan}

\author{Hirotsugu Fujii}
\address{Institute of Physics, University of Tokyo, Tokyo 153-8902, Japan}

\author{Kazunori Itakura}
\address{KEK Theory Center, IPNS,
          High Energy Accelerator Research Organization (KEK), \\
           1-1 Oho, Tsukuba, Ibaraki, 305-0801, Japan}
\address{Department of Particle and Nuclear Studies, \\
Graduate University for Advanced Studies (SOKENDAI), 
1-1 Oho, Tsukuba, Ibaraki 305-0801, Japan}

\author{Yohei Saito}
\address{KEK Theory Center, IPNS,
          High Energy Accelerator Research Organization (KEK) \\
           1-1 Oho, Tsukuba, Ibaraki, 305-0801, Japan}

\begin{abstract}

The dynamic critical exponent $z$ is studied in two different theoretical frameworks:
one is the effective theory of a time-dependent Ginzburg-Landau model, i.e.,
model A in the classification of Hohenberg and Halperin, 
and the other is the microscopic finite-temperature field theory in the imaginary time formalism.
Taking an $O(N)$ scalar model as an example and carrying out the $1/N$ expansion up to the next-to-leading order (NLO) in the one-particle-irreducible (1PI) and two-particle-irreducible (2PI) effective actions, we compare the low-energy and low-momentum (infrared) behavior of the two-point functions in the two theories.
At the NLO of the 1PI $1/N$ expansion the infrared behavior of the two-point functions in the effective and microscopic theories is very much different from each other: 
it is dominated by the diffusive mode with, $z=2+\frac{4}{3\pi^2}\frac{1}{N}$, in model A, while in the microscopic theory it is dominated by the propagating mode with $z=1-\frac{16}{3\pi^2}\frac{1}{N}$ or
$z=2-\frac{32}{3\pi^2}\frac{1}{N}$ depending on whether the kinematics is relativistic or nonrelativistic.
In contrast, at the NLO of the 2PI $1/N$ expansion, we find that
the two theories become equivalent for describing the infrared behavior of the two-point function in the sense that 
the self-consistent equation for the two-point function, the Kadanoff-Baym equation, has exactly the same form both in the microscopic and effective theories.
At this point, the relativistic or nonrelativistic kinematics of the bare two point function
in the microscopic theory becomes irrelevant in the critical dynamics.
This implies that the diffusive mode with $z = 2 + {\cal O}(1/N)$ becomes dominant at low energies and momenta even in the microscopic theory at the NLO of the 2PI $1/N$ expansion,
though we do not explicitly solve the Kadanoff-Baym equation.
We also try to improve the estimate of the dynamic critical exponent 
of model A given in the literature, at the NLO of the strict $1/N$ expansion. 
This calculation can be regarded as an approximation to the 2PI NLO calculation of the dynamic critical exponent not only in the effective theory but also in the microscopic theory.
By incorporating the static 2PI correlations into the two-point function we identify the infrared logarithmic term with respect to energy or momentum in the $1/N$ NLO self-energy, 
 from which we determine the critical exponent, $z$.
The obtained critical exponent is slightly smaller than the previous known result and its $N$ dependence is also milder than the previous one. 
\end{abstract}

\maketitle
\section{Introduction}

Relaxation to the equilibrium state of a system at a critical point becomes extremely slow and shows a universal behavior.
Recently, such dynamic critical phenomena have been receiving much interests in various fields of physics 
from condensed matter physics to cosmology \cite{Tauber:2014,Boyanovsky:2006bf,Calzetta:2008}.
The dynamic critical phenomena have successfully been described by effective theories \cite{Tauber:2014,Hohenberg:1977,Ma:2000,Mazenko:2006},
and have been classified into several subclasses from the static universal classes according to the symmetries of the order parameters and
whether there are couplings to other conserved quantities or not \cite{Hohenberg:1977}.
In these effective theories diffusive motions are assumed for the order parameters at the tree level and then  
nonlinear interactions among the order parameters are included together with the interactions of the order parameters and the conserved quantities.
In principle both the effective and microscopic theories should describe critical phenomena equally well,
if one can take into account contributions relevant to dynamic critical phenomena in each theory.
However, it is known, for instance, that  the $1/N$ expansion in the standard method of the one-particle-irreducible (1PI) effective action leads to different results
for the dynamic critical exponent in the effective and microscopic theories \cite{Halperin:1972,Suzuki:1975,Kondor:1974,Abe:1973,Abe:1974,Suzuki:1974}.
Thus, microscopic understandings of dynamic critical phenomena, in particular the generation of the diffusive mode, have not been achieved and still remain a challenge
\cite{Boyanovsky:2000nt,Boyanovsky:2001pa,Saito:2013}.

The method of the two-particle-irreducible (2PI) effective action \cite{Luttinger:1960,Cornwall:1974,Berges:2010} has recently attracted much attention \cite{Bray:1974zz,Alford:2004jj,Berges:2005,Saito:2011xq}.
In this method, self-energy corrections for the two-point function are first summed up and then the expansion is carried out in terms of the {\it full} two-point function.
This is in contrast to the standard method of the 1PI effective action, where the expansion is in terms of the {\it free} two-point function.
The method of the 2PI effective action provides us with a way of systematic resummation of the perturbative expansion.
Therefore, as was suggested in Ref.~\cite{Saito:2013}, it is expected to take into account the secular effects of collisions in the microscopic theory which are considered to be responsible for the diffusive behavior of the two-point function at low energies and momenta.

According to the dynamic scaling hypothesis \cite{Ferrell:1967zz,Ferrell:1968,Halperin:1967, HALPERIN:1969zza}, the inverse of the retarded two-point function at the critical point, $G(\p,p_0)^{-1}$, has the form
\begin{align}
\label{scaling1}
  G(\p,p_0)^{-1} &= |\p|^{2-\eta}g\left(\frac{p_0}{|\p|^z}\right)
\; ,
\end{align}
where $\eta$ ($z$) is the static (dynamic) critical exponent and $p_0$ ($\p$) is the energy (momentum).
This relation implies that the mode energy scales with the momentum as $p_0 \sim |\p|^z$ at the critical point.

For low $p_0$ ($p$), $G(\p,p_0)^{-1}$ is analytic in $p_0$ ($p$)
\begin{align}
\label{scaling2}
  G(\p,p_0)^{-1} =
  \begin{cases}
    a_0 |\p|^{2-\eta} + a_1 p_0 |\p|^{2-\eta-z} + a_2 p_0^2 |\p|^{2-\eta-2z} + \cdots &(\text{low} \ p_0) \; , \\
    b_1p_0^{(2-\eta)/z} + b_2 |\p| p_0^{(2-\eta-1)/z} + \cdots & (\text{low} \ |\p|) \; ,
  \end{cases}
\end{align}
which constrains the asymptotic behavior of the scaling function, $g(x)$,  to be
\begin{align*}
\label{}
  g(x) = 
  \begin{cases}  
    a_0+ a_1 x + a_2 x^2 + \cdots & (\text{small} \ x)  \; , \\
    b_1 x^{(2-\eta)/z} + b_2 x^{(1-\eta)/z} + \cdots & (\text{large} \ x) \; .
  \end{cases}
\end{align*}
Therefore, while the static critical exponent, $\eta$, can be read off from $G(\p,0)^{-1}$,
the dynamic critical exponent, $z$, can be obtained from either $G(0,p_0)^{-1}$ or $\partial G(\p,p_0)^{-1}/{\partial p_0}|_{p_0=0}$ together with 
the knowledge of $\eta$.

The purpose of the present paper is twofold.
Firstly, we would like to clarify the relation between two descriptions of the dynamic critical phenomena,
i.e.\ in terms of the effective theory and in terms of the microscopic theory, paying special attention to the diffusive mode.
Thereby, we would like to resolve the confusions sometimes seen in the literature.
We employ a simple time-dependent Ginzburg-Landau (TDGL) model \cite{Landau:1954} or model A in the classification of Ref.~\cite{Hohenberg:1977} for the effective theory
and the imaginary-time formalism of the field theory at finite temperature \cite{Matsubara:1955,Fetter:1971,Kapusta:1989,LeBellac:2000} for the microscopic theory.
Taking an $O(N)$ scalar model as an example and carrying out the $1/N$ expansion up to the next-to-leading order (NLO)
in the 1PI and 2PI effective actions,
we compare the low-energy and low-momentum behavior of the response function in the effective theory and of the retarded Green's function in the microscopic theory.
We show that two descriptions are equivalent at the NLO of the 2PI $1/N$ expansion, i.e.\ the self-consistent equation for the two-point function, the Kadanoff-Baym equation,
is exactly the same in the effective and microscopic theories, while two descriptions are quite different at the NLO of the 1PI $1/N$ expansion.
Secondly, we would like to explore the possibility of improving the previous calculation \cite{Halperin:1972} of the dynamic critical exponent at the NLO of  the $1/N$ expansion in model A.
By incorporating the static 2PI correlations into the two-point function, we identify the infrared logarithmic term with respect to energy or momentum, from which we determine the critical exponent, $z$.

The outline of the present paper is as follows.
In Sec.~2 we explain the minimal formalism of the effective and microscopic theories. 
Sec.~3 is devoted to the critical exponents with the 1PI effective action.
After we review the calculation of the static critical exponent, the dynamic critical exponent in the effective theory and also in the nonrelativistic field theory for comparison,
we discuss the dynamic critical exponent in the relativistic field theory.
Then we move on to the critical exponents with the 2PI effective action in Sec.~4.
Again, after reviewing the calculation of the static critical exponent, we discuss the dynamic critical exponent.
In Sec.~5 we present the results of our calculation of the dynamic critical exponent, where we incorporate the static 2PI correlations.
We summarize the paper and provide some discussions in Sec.~6.

\section{Effective and Microscopic Theories}
As an example we consider a system  in a $d$-dimensional space, which is in the symmetric (disordered) phase of the $O(N)$ symmetry with a scalar order parameter.
In the present paper, some quantities, such as two-point function or self-energy, appear both in the effective and microscopic theories.
We express quantities with the superscript $E$ ($M$) in the effective (microscopic) theory,
but without superscript $\to$ omit the superscripts in the relation which holds both in the effective and microscopic theories.

In the TDGL theory, the Ginzburg-Landau Hamiltonian, $H^{E}$, is given in terms of the order parameter $\varphi_a(\x) \ (a=1,...,N)$ as 
\begin{align}
  \label{H^E}
  H^{E}[\varphi] = \int d^d\x
    &\left[ \frac{1}{2}\partial_i \varphi_a(\x) \partial_i \varphi_a(\x)
        +\frac{r^E}{2}\varphi_a(\x)\varphi_a(\x)
        +\frac{u^E}{4! N}(\varphi_a(\x)\varphi_a(\x))^2 
        - h_a(\x)\varphi_a(\x)\right],
\end{align}
where $h_a(\x)$ is the external field. 
Consider a process in which the system slightly out of equilibrium relaxes to equilibrium.
In the approximation we consider in the present paper (i.e., next-leading-order in the $1/N$ expansion), it is sufficient to consider the dynamics of the order parameter since the coupling to an $O(N)$ charge appears only in the higher orders. 
We also assume that the coupling to the energy is unimportant. Thus, the dynamics is categorized as model A.
The time dependence of the order parameter for such a relaxation process is described by the Langevin equation \cite{Landau:1954}
\begin{align}
  \label{Langevin}
  \frac{\partial\varphi_a(\x,t)}{\partial t} = - \Gamma_0\frac{\delta H^{E}[\varphi]}{\delta \varphi_a(\x,t)} + \zeta_a(\x,t).
\end{align}
$\Gamma_0$ is the relaxation constant and $\zeta_a(\x,t)$ is the noise, whose average and correlation at temperature $T$ are assumed to be given respectively as
\begin{align*}
  \langle\zeta_a(\x,t)\rangle &= 0, \\
  \langle\zeta_a(\x,t)\zeta_b(\x',t')\rangle &= 2T\Gamma_0\delta^d(\x-\x')\delta(t-t')\delta_{ab}.
\end{align*}
The Fourier transformation of Eq.~(\ref{Langevin})  yields
\begin{align*}
  -i p_0 \varphi_a(\p,p_0) = &- \Gamma_0 \left[ (\p^2 + r^E) \varphi_a(\p,p_0)
  + \frac{u^E}{3! N} [\varphi^3]_a(\p,p_0) - h_a(\p,p_0) \right] + \zeta_a(\p,p_0)
\; , 
\end{align*}
where\footnote{
In this paper we use the abbreviated notation $\displaystyle{\int_{{\boldsymbol k},k_0} = \int \frac{d^3{\boldsymbol k}}{(2\pi)^3}\frac{dk_0}{2\pi}}$
and $\displaystyle{\int_{{\boldsymbol k}} = \int \frac{d^3{\boldsymbol k}}{(2\pi)^3}}$.
}

\begin{align*}
  [\varphi^3]_a(\p,p_0) = \int_{\q,q_0} \int_{{\boldsymbol k},k_0} \varphi_a(\p-\q,p_0-q_0) \varphi_b(\q-{\boldsymbol k},q_0-k_0) \varphi_b({\boldsymbol k},k_0).
\end{align*}
The response function, $G^{E}_{ab}(\p,p_0)$, and the correlation function, $C^{E}_{ab}(\p,p_0)$, are respectively defined by
\begin{align*}
  G^{E}_{ab}(\p,p_0) &= \lim_{h \rightarrow 0} \frac{\langle \varphi_a(\p,p_0)\rangle_h - \langle \varphi_a(\p,p_0)\rangle_{h=0}}{h_b(\p,p_0)}, \\
  C^{E}_{ab}(\p,p_0) &=  \int dt d^d\x {\rm e}^{i(p_0t-\p\cdot\x)} \langle \varphi_a(\x,t) \varphi_b(\0,0) \rangle_{h=0},
\end{align*}
where $\langle \cdots \rangle_h$ denotes the average over the noise with the presence of the external field, $h$.
They satisfy the following relation (fluctuation-dissipation theorem)
\begin{align*}
  C^{E}_{ab}(\p,p_0) =  \frac{2T}{p_0}{\rm Im}G^{E}_{ab}(\p,p_0).
\end{align*}

In the imaginary-time formalism of finite temperature field theory
\cite{Matsubara:1955,Fetter:1971,Kapusta:1989, LeBellac:2000},
the Hamiltonian of the relativistic theory is given in terms of the microscopic field, $\hat\varphi_a(\x,t)$, and its conjugate momentum, $\hat \pi_a(\x,t)$, as
\begin{align}
  \label{H^M}
  \hat H^{M}[\varphi]=\int d^d\x
  \left[  \frac{1}{2}\hat \pi_a(\x,t) \hat \pi_a(\x,t) + \frac{1}{2}\partial_i \hat \varphi_a(\x,t) \partial_i \hat \varphi_a(\x,t)
        +\frac{r^M}{2}\hat \varphi_a(\x,t)\hat \varphi_a(\x,t) + \frac{u^M}{4!N}(\hat \varphi_a(\x,t)\hat \varphi_a(\x,t))^2\right].
\end{align}
(The Hamiltonian of the nonrelativistic theory is given similarly, e.g.\ \cite{Fetter:1971}.)
The effective Hamiltonian, Eq.~(\ref{H^E}), is of the same form as the microscopic Hamiltonian, Eq.~(\ref{H^M}), except for the kinetic energy term
and the term with the external filed.
If the former is regarded to be derived from the latter by some coarse graining or renormalization procedures, the terms of the former should include
contributions of fluctuations of high frequency modes of the latter.
Therefore, in general the parameters of the former, $r^E$ and $u^E$, are different from those of the latter, $r^M$ and $u^M$,
which is the reason we distinguish them. 
The imaginary-time Green's function is defined by
\begin{align*}
  G^{M}_{ab}(\p,i\omega_n) = \int _0^{\beta} d\tau \int d^d\x \ {\rm e}^{i(\omega_n\tau-\p\cdot\x)}
  {\rm Tr} \left\{{\rm e}^{-\beta \hat H^{M}} 
  \left[\theta(\tau)\hat\varphi_a(\x,-i\tau) \hat\varphi_b(\0,0) + \theta(-\tau) \hat\varphi_b(\0,0)\hat\varphi_a(\x,-i\tau)\right]\right\},
\end{align*}
where $\omega_n$ is the Matsubara frequency \cite{Matsubara:1955} and Tr represents the sum over a complete set of states in the Hilbert space.
Then, the real-time retarded Green's function,
\begin{align*}
  G^{M}_{ab}(\p,p_0) = \int_{-\infty}^{\infty} dt \int d^d\x \ {\rm e}^{i(p_0t-\p\cdot\x)} {\rm Tr} \left\{{\rm e}^{-\beta \hat H^{M}}
  \theta(t)\left[\hat\varphi_a(\x,t) \hat\varphi_b(\0,0) - \hat\varphi_b(\0,0)\hat\varphi_a(\x,t)\right]\right\},
\end{align*}
is obtained by the analytic continuation of the imaginary-time Green's function as $i\omega_n \rightarrow p_0+ i\epsilon$
at the end of the calculation.

\section{1PI 1/N expansion in effective and microscopic theories}
Consider the Schwinger-Dyson equation in the effective and microscopic theories,
\begin{align}
\label{SD}
  G(\p,p_0)^{-1} = G_0(\p,p_0)^{-1} + \Sigma(\p,p_0).
\end{align}
$G(\p,p_0)$ and $G_0(\p,p_0)$ are respectively the full and bare two-point functions,
i.e.\ response (retarded Green's) functions in the effective (microscopic) theory, and $\Sigma(\p,p_0)$ is the self-energy.
$G_0(\p,p_0)^{-1}$ is given in the effective theory as
\begin{align}
  \label{G_0_eff}
  G^{E}_{0}(\p,p_0)^{-1} = \p^2 -i p_0 \Gamma_0^{-1},
\end{align}
while in the microscopic theory as
\begin{align}
  \label{G_0_micro}
  G^{M}_{0}(\p,p_0)^{-1} = 
  \begin{cases}
    \displaystyle{\frac{\p^2}{2M}} - p_0 & \text{(nonrelativistic)} \\
    \p^2 - p_0^2 & \text{(relativistic)}
  \end{cases}
\end{align}
at the critical point.
\footnote{The term in the self energy, which does not depend on the energy and the momentum but depends on the temperature, is taken into account
as the temperature-dependent mass in the bare two-point function.
This temperature dependent mass vanishes at the critical point.}
The dispersion relation is given by $p_0 = -i \Gamma_0 \p^2$ in the effective theory and by $p_0^2 = \p^2$ ($p_0 = \frac{\p^2}{2M}$)
in the relativistic (nonrelativistic) microscopic theory.
The former is called the diffusive mode and the latter the propagating mode.

In both theories the self-energy is given in the $1/N$ expansion as
\begin{align*}
  \Sigma(\p,p_0) &= \Sigma_{\rm LO}(\p,p_0) + \Sigma_{\rm NLO}(\p,p_0) + \cdots\\
  &=
  \raisebox{-3mm}{\includegraphics[width=1cm]{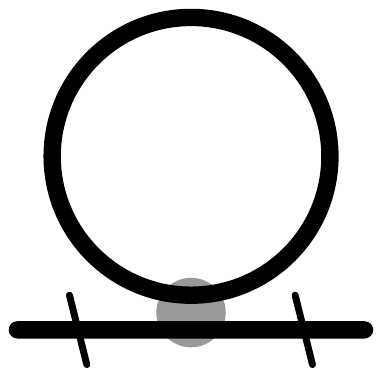}}
  \;
  +
  \;
  \raisebox{-3mm}{\includegraphics[width=2cm]{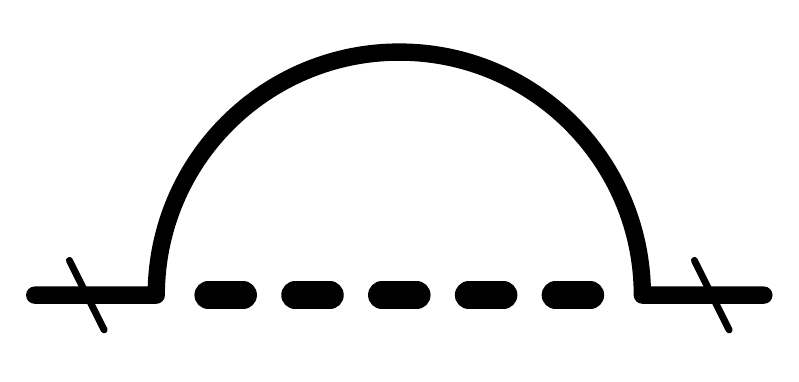}}
  \;
  +
  \;
  \cdots,
\end{align*}
where the external lines are amputated,
$\Sigma_{\rm LO}(\p,p_0)$ and $\Sigma_{\rm NLO}(\p,p_0)$ are respectively the self-energies at the leading order (LO) 
and the NLO of the $1/N$ expansion ($\Sigma_{\rm LO}(\p,p_0)={\cal O}(1)$ and $\Sigma_{\rm NLO}(\p,p_0)={\cal O}(1/N)$),
and the dashed line denotes the sum of bubble diagrams
\begin{align*}
\raisebox{-3mm}{\includegraphics[width=2cm]{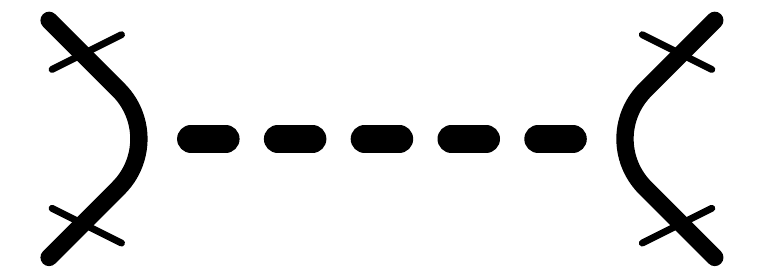}}
\;
   =
\;
\raisebox{-4mm}{\includegraphics[width=10mm]{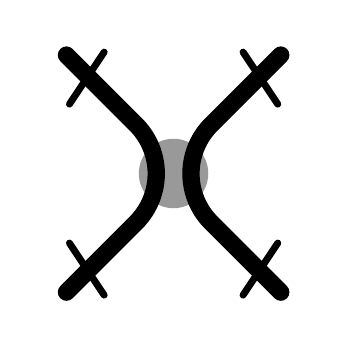}}
\;
 + \cdots +
\;
\raisebox{-3mm}{\includegraphics[width=2.5cm]{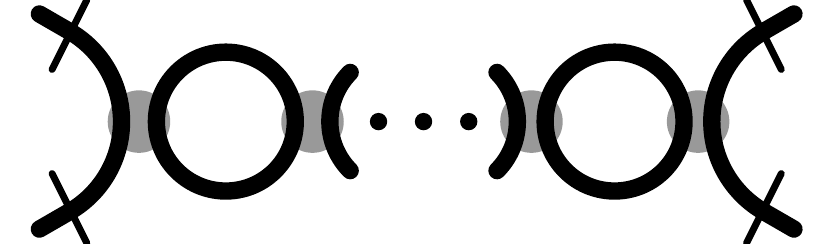}}
\;
\ \ \ +\ \ \cdots.
\end{align*}
As can be seen in the above, bubble diagrams appear in the calculation of the NLO self-energy.
In Appendix~A we derive the expressions for general bubble diagrams in the effective and microscopic theories,
which will be used below.

We define the function $f(p_0)$ by
\begin{align*}
  f^E(p_0) &= \frac{T}{p_0}, \\
  f^M(p_0) &= \frac{1}{\exp(p_0/T)-1},
\end{align*}
where $f^M(p_0)$ coincides with the Bose distribution function when $p_0>0$.
In terms of $f(p_0)$  we can write $\Sigma_{\rm LO}(\p,p_0)$ and $\Sigma_{\rm NLO}(\p,p_0)$ in the same form as
\begin{align}
  \label{Sigma_LO}
  \Sigma_{\rm LO}(\p,p_0) = \text{(vacuum term)} + \frac{u}{6}\int_{\q,q_0} f(q_0){\rm Im}G_0(\q,q_0)
\end{align}
and
\begin{align*}
  \label{}
  \Sigma_{\rm NLO}(\p,p_0) =  \frac{u}{3N}\int_{\q,q_0} & \left\{ G_0(\p-\q,p_0-q_0) f(q_0){\rm Im}D_0(\q,q_0) \right. \nonumber\\
  & \ \left. - f(q_0-p_0){\rm Im}G_0(\p-\q,p_0-q_0) D_0(\q,q_0) \right\},
\end{align*}
where
\begin{align}
  \label{D_0}
  D_0(\p,p_0) = \frac{1}{1+\displaystyle{\frac{u}{6}}\Pi_0(\p,p_0)} 
\end{align}
and
\begin{align}
  \label{Pi_0}
  \Pi_0(\p,p_0) &=2 \int_{\q,q_0} G_0(\p-\q,p_0-q_0)\left(f(q_0)-f(-q_0)\right){\rm Im}G_0(\q,q_0).
\end{align}
In Eq.~(\ref{Sigma_LO}) the vacuum term exists only in the microscopic theory.
Also, $\Sigma_{\rm LO}(\p,p_0)$ does not depend on $\p$ and $p_0$, which will be simply written as $\Sigma_{\rm LO}$ hereafter.

As far as the critical behavior is concerned, $f^M(p_0)=[\exp(p_0/T)-1]^{-1}$ in the microscopic theory can be replaced by its high temperature expansion,
$f^M(p_0) \sim T/p_0 \ (=f^E(p_0))$, which will be used from now on.
The vacuum term in $\Sigma^{M}_{\rm LO}$ is also irrelevant since it is renormalized into the mass.
Then, the only difference in the effective and microscopic theories is in the form of the bare two-point function, $G_0(\p,p_0)$.

As is shown in Appendix B, one can separate the static and dynamic parts of $\Sigma_{\rm NLO}(\p,p_0)$ as
\begin{align}
\label{Sigma_NLO}
  \Sigma_{\rm NLO}(\p,p_0) &= \frac{u T}{3N}\int_{\q} G_0(\p-\q,0) D_0(\q,0) \nonumber\\
  &\phantom{=} -\frac{2u T}{3N}p_0\int_{\q,q_0}
  {\rm Im}G_0(\p-\q,p_0-q_0) \frac{1}{q_0(p_0-q_0)}
  \left\{ D_0(\q,0) - D_0(\q,q_0)\right\} \nonumber\\
  &\equiv \Sigma_{\rm NLO}(\p,0) + \Delta \Sigma_{\rm NLO}(\p,p_0)
\end{align}
and similarly of $\Pi_0(\p,p_0)$ as
\begin{align}
\label{Pi}
  \Pi_0(\p,p_0) &= T \int_{\q} G_0(\p-\q,0) G_0(\q,0) \nonumber\\
  &\phantom{=} -2T p_0 \int_{\q,q_0}
  {\rm Im}G_0(\p-\q,p_0-q_0) \frac{1}{q_0(p_0-q_0)}
  \left\{ G_0(\q,0) - G_0(\q,q_0)\right\} \nonumber\\
  &\equiv \Pi_0(\p,0) + \Delta \Pi_0(\p,p_0).
\end{align}
Clearly, $\Sigma_{\rm NLO}(\p,0)$ coincides with that of the static theory if $u T$ is replaced by $u_{\rm static}$.
\subsection{Static critical exponent}
The static part of the Schwinger-Dyson equation, Eq.~(\ref{SD}), reads
\begin{align*}
  G(\p,0)^{-1} &= G_0(\p,0)^{-1} + \Sigma(\p,0).
\end{align*}
As is seen in Eqs.~(\ref{G_0_eff}) and (\ref{G_0_micro}), $G_0(\p,0)^{-1}$ is the same in the effective and microscopic theories, $G_0(\p,0)^{-1} = \p^2$.
(In the nonrelativistic microscopic theory $G^{M}_0(\p,0)^{-1} = \p^2/2M$
but the factor $1/2M$ can be trivially rescaled.)
Thus, the effective and microscopic theories have exactly the same static part.

At the critical point $G(\p,0)^{-1}$ has the form with a scale $\Lambda$,
\begin{align*}
\label{}
    G(\p,0)^{-1} &= \Lambda^\eta |\p|^{2-\eta} \\
    &= \p^2 - \eta \p^2\log(|\p|/\Lambda) + {\cal O}(\eta^2).
\end{align*}
In Ref.~\cite{Ma:1973} it was shown that the static self-energy at the NLO of the $1/N$ expansion,
$\Sigma_{\rm NLO}(\p,0)$, behaves for low momentum, $|\p|$, in the $d$-dimensional space as
\begin{align*}
  [\Sigma_{\rm NLO}(\p,0)]_{\log |\p|} = - 4\left(\frac{4}{d}-1\right)\frac{\sin\pi(\frac{1}{2}d-1)}{\pi(\frac{1}{2}d-1)B(\frac{1}{2}d-1,\frac{1}{2}d-1)}\frac{1}{N} \p^2\log |\p|,
\end{align*}
where $B(a,b)$ is the Beta function.
Then, the static critical exponent, $\eta$, is obtained as
\begin{align*}
  \eta = [\Sigma_{\rm NLO}(\p,0)]_{\log |\p|} = - 4\left(\frac{4}{d}-1\right)\frac{\sin\pi(\frac{1}{2}d-1)}{\pi(\frac{1}{2}d-1)B(\frac{1}{2}d-1,\frac{1}{2}d-1)}\frac{1}{N}.
\end{align*}
In particular for $d=3$, this becomes
\begin{align*}
  \eta = \frac{8}{3\pi^2}\frac{1}{N}.
\end{align*}
This result is common for the effective and microscopic theories and irrespective of whether the system is non-relativistic or relativistic.
Also, the result is independent of the coupling constant,
which can be understood as universality is realized by summing over terms of different orders of the coupling constant, $u$,
but with the same order of $N$ \cite{Ma:2000}.
\subsection{Dynamic critical exponent}
%
%
In the TDGL theory \cite{Landau:1954,Ma:2000,Mazenko:2006}, the bare response function is given by Eq.~(\ref{G_0_eff}),
which describes the diffusive mode.
This means $z=2$ at the tree level.

If one takes $\p = 0$, the Schwinger-Dyson equation, Eq.~(\ref{SD}), becomes
\begin{align*}
  G^{E}(0,p_0)^{-1} &= G_0^{E}(0,p_0)^{-1} + \Sigma^{E}(0,p_0)
  = -i\Gamma_0^{-1} p_0 + \Sigma^{E}(0,p_0).
\end{align*}
At the critical point, we introduce the anomalous dimension $\lambda$ as 
a deviation from the power of $p_0$ in $G^E_0(0,p_0)^{-1}$.
Since in the effective theory there is no distinction due to the kinematics, either relativistic or nonrelativistic, (see Eq.~(\ref{G_0_eff})), one finds
\begin{align*}
\label{}
  G^{E}(0,p_0)^{-1} = -i\Gamma_0^{-1} \Lambda^{\lambda} p_0^{1-\lambda} = -i\Gamma_0^{-1}\left\{p_0 - \lambda p_0 \log(p_0/\Lambda)\right\} + {\cal O}(\lambda^2).
\end{align*}
Then the dynamic critical exponent, $z$, is obtained from $\lambda$ and the static critical exponents, $\eta$, as $z=(2-\eta)/(1-\lambda)$ (see Eq.~(\ref{scaling2})).

At the NLO of the $1/N$ expansion, $\lambda$ is extracted from the dynamic part of the self-energy, as $\eta$ is from the static part,
\begin{align*}
    [\Delta\Sigma^{E}_{\rm NLO}(0,p_0)]_{\log p_0} = i \Gamma_0^{-1}\lambda p_0 \log p_0.
\end{align*}
In Ref.~\cite{Halperin:1972} $\lambda$ was obtained as
\begin{align*}
    \lambda &= \frac{1}{B(2-\frac{d}{2},\frac{d}{2})}
    \left\{\frac{1}{\int_0^{1/2}dx[x(2-x)]^{d/2-2}}-\frac{1}{\frac{1}{2}B(\frac{1}{2}d-1,\frac{1}{2}d-1)}\right\}\frac{1}{N} \\
    &= \frac{1}{4}\frac{d}{4-d}\left\{\frac{B(\frac{1}{2}d-1,\frac{1}{2}d-1)}{\int_0^{1/2}dx[x(2-x)]^{d/2-2}}-2\right\}\eta.
\end{align*}
Then, $z$ is obtained as
\begin{align*}
    z = \frac{2-\eta}{1-\lambda} = 2-\eta+2\lambda = 2 +c\eta,
\end{align*}
where
\begin{align*}
   c = \frac{4}{4-d}\left\{\frac{dB(\frac{1}{2}d-1,\frac{1}{2}d-1)}{8\int_0^{1/2}dx[x(2-x)]^{d/2-2}}-1\right\}
\end{align*}
and $O\left(\displaystyle{1/N^2}\right)$ terms are neglected.
The same result was obtained also in Ref.~\cite{Suzuki:1975}.
In particular for $d=3$, these become 
\begin{align*}
   \lambda = \frac{2}{\pi^2}\frac{1}{N}, \quad
   z = \frac{2-\eta}{1-\lambda} = \frac{{2-\frac{8}{3\pi^2}\frac{1}{N}}}{{1-\frac{2}{\pi^2}\frac{1}{N}}}
    = 2 + \frac{4}{3\pi^2}\frac{1}{N}, \ \quad
   c = \frac{1}{2}.
\end{align*}

%
%
In the finite-temperature field theory the bare retarded Green's function is given by Eq.~(\ref{G_0_micro})
which describes the propagating mode.
This means $z=2$ for the nonrelativistic theory while $z=1$ for the relativistic theory at the tree level.

Similarly as discussed in the effective theory, we again define the 
anomalous dimension $\lambda$ as a deviation from the power in $p_0$ of 
$G_0^M(0,p_0)^{-1}$. This time, we need to distinguish relativistic and non-relativistic cases (see Eq.~(\ref{G_0_micro})). Therefore, at the critical point $G^{M}(0,p_0)^{-1}$ has the form
\begin{align*}
\label{}
  G^{M}(0,p_0)^{-1} =
  \begin{cases}
    - \Lambda^\lambda p_0^{1-\lambda} = - p_0 + p_0 \lambda \log (p^0/\Lambda) + O(\lambda^2) & \text{(nonrelativistic)} \\
    - \Lambda^\lambda p_0^{2-\lambda} = - p_0^2 + p_0^2 \lambda \log (p^0/\Lambda) + O(\lambda^2) & \text{(relativistic)}
  \end{cases}.
\end{align*}

In the nonrelativistic theory the dynamic critical exponent is obtained at the NLO of the $1/N$ expansion
by Kondor and Sz\'epfalusy \cite{Kondor:1974}, by Abe, Hikami \cite{Abe:1973,Abe:1974} and by Suzuki and Tanaka \cite{Suzuki:1974}.
The obtained dynamic critical exponent, $z$, depends on the space dimension, $d$, in a rather complicated way.
Here, we concentrate on the case, $d=3$.
Then, $\Delta\Sigma_{\rm NLO}(0,p_0)$ has a real logarithmic term
\begin{align*}
  [\Delta\Sigma^{M}_{\rm NLO}(0,p_0)]_{\log p_0} = -\frac{4}{\pi^2} \frac{1}{N} p_0 \log p_0 \qquad (p_0 > 0)
\end{align*}
and neglecting ${\cal O}(1/N^2)$ terms, we find 
\begin{align*}
  z=\frac{2-\eta}{1-\lambda}=\frac{{2-\frac{8}{3\pi^2}\frac{1}{N}}}{{1+\frac{4}{\pi^2}\frac{1}{N}}}=2-\frac{32}{3\pi^2}\frac{1}{N}.
\end{align*}

In the relativistic theory, we calculated the self-energy, $\Delta\Sigma^{M}_{\rm NLO}(0,p_0)$ at the NLO of the $1/N$ expansion.
The details of the calculation are shown in Appendix C.
We found a real logarithmic term
\begin{align*}
  [\Delta\Sigma^{M}_{\rm NLO}(0,p_0)]_{\log p_0} = -\frac{8}{\pi^2} \frac{1}{N} p_0^2 \log p_0 \qquad (p_0 > 0)
\end{align*}
and  neglecting ${\cal O}(1/N^2)$ terms, we obtain  
\begin{align*}
  z=\frac{2-\eta}{2-\lambda}=\frac{{2-\frac{8}{3\pi^2}\frac{1}{N}}}{{2+\frac{8}{\pi^2}\frac{1}{N}}}=1-\frac{16}{3\pi^2}\frac{1}{N}.
\end{align*}
It is interesting that in the relativistic theory both the LO and NLO contributions to $z$ are twice as large as those in the nonrelativistic theory.\footnote{
The same results were obtained by J.~Berges and his collaborator.
We would like to thank J.~Berges for informing us about their results.}

It should be noted that in the effective theory the obtained $z$ is for the imaginary diffusive mode,
while it is for the real propagating mode in the microscopic theory at the NLO of the $1/N$ expansion.
If the diffusive mode were dynamically generated in the microscopic theory, it would have the critical exponent, $z \sim 2$, and dominate over the propagating mode at low energies and momenta.
Therefore, it is important if and how the diffusive mode is generated in the microscopic theory.\footnote{
In Ref.~\cite{Saito:2013} if the diffusive mode appears at the NLO of the 1/N 1PI and 2PI in the two-point function was studied.}
As is shown in Appendix C, the imaginary part of the NLO self-energy, ${\rm Im}\Sigma^{M}_{\rm NLO}(0,p_0)$, behaves as
\begin{align*}
  {\rm Im}\Sigma^{M}_{\rm NLO}(0,p_0) \propto \frac{1}{N} p_0^2 \frac{1}{\log p_0} \qquad (p_0 > 0),
\end{align*}
which is smaller than the real part.
Thus, at the NLO of the $1/N$ expansion the imaginary part of the self-energy exists, but it is not sufficiently large at low energies and momenta to generate the diffusive mode. 

\section{2PI 1/N expansion in effective and microscopic theories}
In the method of the 2PI effective action, self-energy corrections for the two-point function are first summed up and then the expansion is carried out in terms of the full two-point function, $G$.
Then, the self-energy is given as a functional of $G$ and the Schwinger-Dyson equation, Eq.~(\ref{SD}), is formally changed to 
\begin{align}
  \label{SD_2PI}
  G(\p,p_0)^{-1} &= G_0(\p,p_0) ^{-1}+ \Sigma[G](\p,p_0).
\end{align}
This equation is regarded as a self-consistent equation for the full two-point function, $G$,
and  is called the Kadanoff-Baym equation \cite{Kadanoff:1961,Baym:1962,Kadanoff:1962}.
It is written in the 2PI $1/N$ expansion up to the NLO as
\begin{align}
  \label{SD_2PI'}
  G(\p,p_0)^{-1} &= G_0(\p,p_0) ^{-1}+ \Sigma_{\rm LO}[G] + \Sigma_{\rm NLO}[G](\p,p_0),
\end{align}
in which $\Sigma_{\rm LO}[G]$ and $\Sigma_{\rm NLO}[G]$ are given by Eqs.~(\ref{Sigma_LO}) and (\ref{Sigma_NLO}), respectively,
but with the bare two-point function, $G_0$, replaced by the full two-point function, $G$. 
At the critical point the inverse of the bare and full two-point functions at zero energy and momentum vanish,
i.e.\ $G_0(\p=0,p_0=0)^{-1}=G(\p=0,p_0=0)^{-1}=0$.
Using these and subtracting from Eq.~(\ref{SD_2PI}) the same expression for zero energy and momentum we can write
\begin{align*}
\label{}
  G(\p,p_0)^{-1} &= G_0(\p,p_0) ^{-1}+ \left\{\Sigma_{\rm NLO}[G](\p,p_0)-\Sigma_{\rm NLO}[G](0,0)\right\}.
\end{align*}
The static and dynamic parts of this equation, respectively, become
\begin{align}
  \label{KB_static}
  G(\p,0) ^{-1}= G_0(\p,0) ^{-1}+ \left\{\Sigma_{\rm NLO}[G](\p,0)-\Sigma_{\rm NLO}[G](0,0)\right\}
\end{align}
and
\begin{align}
  \label{KB_dynamic}
  \Delta G(\p,p_0)^{-1} =  \Delta G_0(\p,p_0)^{-1} + \Delta\Sigma_{\rm NLO}[G](\p,p_0),
\end{align}
where $G(\p,p_0) ^{-1}= G(\p,0)^{-1} + \Delta G(\p,p_0)^{-1}$ and 
$G_0(\p,p_0)^{-1} = G_0(\p,0)^{-1} + \Delta G_0(\p,p_0)^{-1}$.

\subsection{Static critical exponent}
Substituting Eq.~(\ref{Sigma_NLO}) and then (\ref{D_0}) into Eq.~(\ref{KB_static}), the static part becomes
\begin{align*}
  G(\p,0)^{-1} &= G_0(\p,0)^{-1} + \frac{uT}{3N}\int_{\q} \left[G(\p-\q,0) - G(-\q,0)\right] D(\q,0) \\
  &= G_0(\p,0)^{-1} + \frac{uT}{3N}\int_{\q} \left[G(\p-\q,0) - G(-\q,0)\right] \left[1+\frac{u T}{6}\int_{{\boldsymbol k}} G(\q-{\boldsymbol k},0) G({\boldsymbol k},0)\right]^{-1}.
\end{align*}
This equation involves only the static part of the two-point function, $G(\p,0)^{-1}$, and coincides  with that of Ref.~\cite{Alford:2004jj}
when one makes the replacement $u T \rightarrow u_{static}$ and $G(\p,0) \rightarrow G_{static}(\p)$ and notices $G_0(\p,0)^{-1}=\p^2$.
One determines the static critical exponent $\eta$ by comparing the leading terms of the left and right hand side of this equation for low momentum
where $\left[1+\frac{u T}{6}\int_{{\boldsymbol k}} G(\q-{\boldsymbol k},0) G({\boldsymbol k},0)\right]^{-1}$ can be replaced by $\frac{6}{u T}\left[\int_{{\boldsymbol k}} G(\q-{\boldsymbol k},0) G({\boldsymbol k},0)\right]^{-1}$.
If $\eta > 0$ and $G(\p,0)^{-1} \propto |\p|^{2-\eta}$, the term, $G_0(\p,0)^{-1} = \p^2$, is subleading and one has 
\begin{align}
  G(\p,0)^{-1} &= \frac{2}{N}\int_{\q} \left[G(\p-\q,0) - G(-\q,0)\right] \left[\int_{{\boldsymbol k}} G(\q-{\boldsymbol k},0) G({\boldsymbol k},0)\right]^{-1} + (\text{subleading terms}).
\end{align}
This is the static part of the Kadanoff-Baym equation at the critical point.
This equation is universal in the sense that it is independent of the coupling constant $u$ and the temperature.
By substituting $G(\p,0)^{-1} \propto |\p|^{2-\eta}$ into this equation and comparing the coefficients of the leading terms, one has
\begin{align}
\label{eta_2PI}
    \eta = \frac{4}{N} \left(\frac{4-d-2\eta}{2-\eta}\right) \frac{\Gamma(1+\frac{1}{2}\eta)\Gamma(1-\frac{1}{2}\eta)\Gamma(2-\eta)\Gamma(d+\eta-2)}
    {\Gamma(\eta+\frac{1}{2}d-1)\Gamma(1+\frac{1}{2}d-\frac{1}{2}\eta)\Gamma(2-\eta-\frac{1}{2}d)\Gamma(\frac{1}{2}d+\frac{1}{2}\eta-1)},
\end{align}
whose solution is the static critical exponent, $\eta$, at the NLO 2PI $1/N$ expansion \cite{Bray:1974zz,Alford:2004jj}.

Figure~1 shows the resulting $\eta$ for $d=3$, where the result of the NLO 1PI $1/N$ expansion is also shown for comparison. 

\begin{figure}[th]
  {\includegraphics[height=5cm]{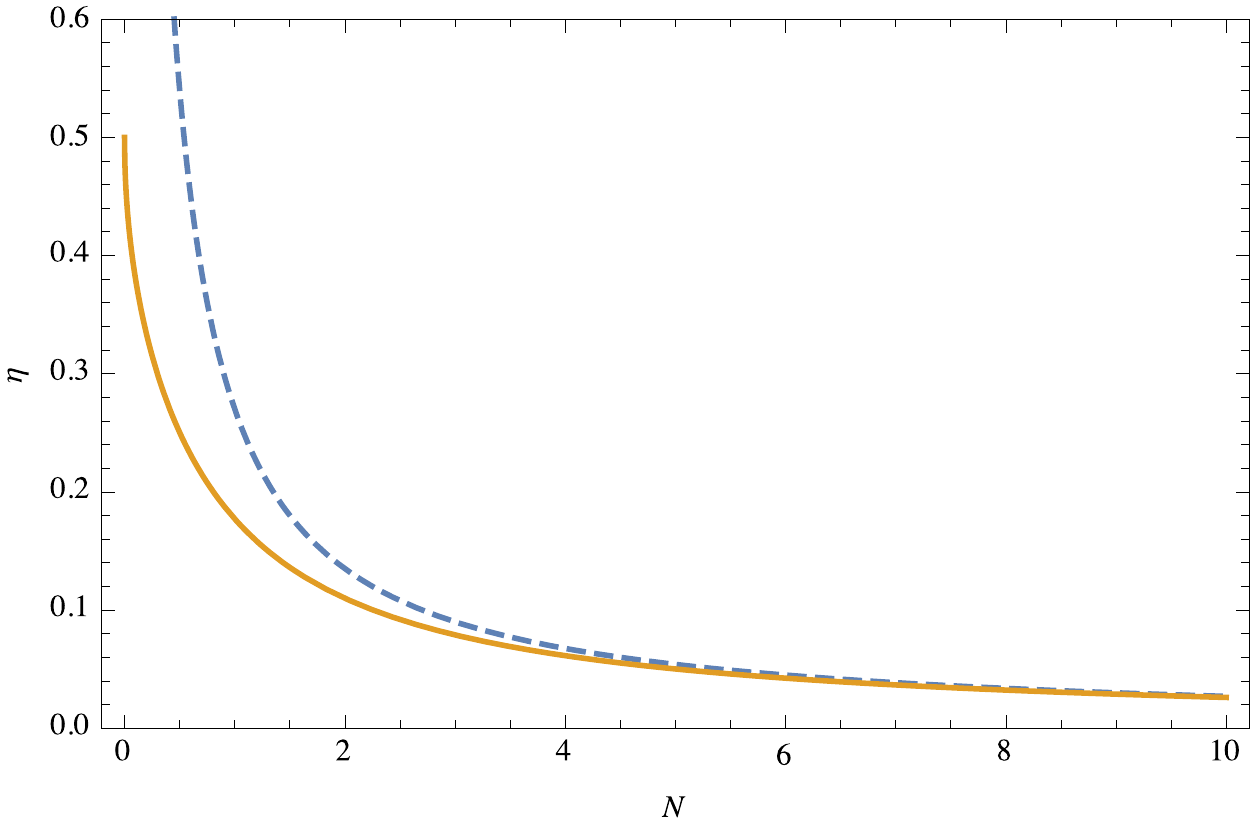}}
\caption{Static critical exponent $\eta$ vs.\ $N$ for $d=3$. The solid curve is the result of the NLO 2PI $1/N$ expansion \cite{Bray:1974zz,Alford:2004jj} and the dashed curve is that of the 1PI $1/N$ expansion \cite{Halperin:1972,Suzuki:1975}, $\eta = \frac{8}{3\pi^2}\frac{1}{N}$.}
\label{eta}
\end{figure}

\subsection{Dynamic critical exponent}
As we have noticed in Sec.~3, the NLO self-energy is exactly the same in the effective and microscopic theories, which is a functional of the response function
in the former and of the retarded Green's function in the latter.
Then, the dynamic part of the Kadanoff-Baym equations in the two theories are different only in the first term of the right-hand-side of Eq.~(\ref{KB_dynamic}), $\Delta G_0(\p,p_0)^{-1}$,
which is $-i p_0 \Gamma_0^{-1}$ in the effective theory and $- p_0^2 \ (- p_0)$ in the relativistic (nonrelativistic) microscopic theory.  
If the Kadanoff-Baym equation in the effective theory has a nontrivial scaling solution at low energies and momenta with $z=2+{\cal O}(1/N)$, as we expect,
$\Delta G_0^E(\p,p_0)^{-1} (= -i p_0 \Gamma_0^{-1})$ will become subleading in Eq.~(\ref{KB_dynamic}).
Then, also in the relativistic (nonrelativistic) microscopic theory $\Delta G_0^M(\p,p_0)^{-1} (= - p_0^2 \ (- p_0))$ will become
subleading and the Kadanoff-Baym equation should have the same solution as in the effective theory.
Namely, {\it effective and microscopic theories are equivalent in the NLO 2PI $1/N$ expansion as far as the critical behavior is concerned.
It should be also noted whether the microscopic theory is relativistic or nonrelativistic is irrelevant as well for the scaling solution in the NLO 2PI $1/N$ expansion.}
This is in clear contrast to the NLO 1PI $1/N$ expansion where the modes of the two theories at low energies and momenta are very much different as we have seen.

Then, the dynamic part of the Kadanoff-Baym equation becomes
\begin{align*}
  \Delta G(\p,p_0)^{-1}
  &= -\frac{2uT}{3N}p_0\int_{\q,q_0}
  {\rm Im}G(\p-\q,p_0-q_0) \frac{1}{q_0(p_0-q_0)}
  \left\{ D(\q,0) - D(\q,q_0)\right\} \\
  &= -\frac{4T}{N}p_0\int_{\q,q_0} {\rm Im}G(\p-\q,p_0-q_0) \frac{1}{q_0(p_0-q_0)} \left\{ \Pi(\q,0)^{-1} - \Pi(\q,q_0)^{-1}\right\}
\end{align*}
where $D(\q,q_0) = \left[1+\frac{u}{6}\Pi(\q,q_0)\right]^{-1}$ is replaced by $\frac{6}{u}\Pi(\q,q_0)^{-1}$ for low energies and momenta.
By substituting Eqs.~(\ref{Pi_0}) and (\ref{Pi}) into the above we finally obtain the dynamic part of the Kadanoff-Baym equation at the critical point,
\begin{align}
  \label{KB_final}
  \Delta G(\p,p_0)^{-1}
  &= -\frac{4}{N}p_0\int_{\q,q_0}
  {\rm Im}G(\p-\q,p_0-q_0) \frac{1}{q_0(p_0-q_0)} \nonumber \\
  & \phantom{= \frac{2}{N}p_0}
  \times \left\{ \left[\int_{{\boldsymbol k}} G(\q-{\boldsymbol k},0) G({\boldsymbol k},0)\right]^{-1} - \left[4\int_{{\boldsymbol k}, k_0}G(\q-{\boldsymbol k},q_0-k_0) \frac{1}{k_0} {\rm Im}G({\boldsymbol k},k_0)\right]^{-1}\right\} \nonumber \\
  &\phantom{=} + (\text{subleading terms}).
\end{align}
Just like the static part, {\it this equation is independent of the coupling constant and the critical temperature, i.e.\ universality holds.}
In contrast to the static part, however, this equation involves both the dynamic and static parts of the two-point function,  $\Delta G(\p,p_0)^{-1}$ and $G(\p,0)^{-1}$.

The observation that effective and microscopic theories are equivalent in the NLO 2PI $1/N$ expansion and Eq.~(\ref{KB_final}) is one of the main results of the present paper.
While only the static critical exponent, $\eta$, has to be determined to solve the static part of the Kadanoff-Baym equation,
the scaling function, $g$ in Eq.~(\ref{scaling1}), has to be determined in addition to the dynamic critical exponent, $z$, to solve the dynamic part.
This makes it far more difficult to determine the dynamic scaling behavior than the static one. 
We are now trying to solve the dynamic part of the Kadanoff-Baym equation and hope to report the results in a near future.

\section{improvement of the calculation of dynamic critical exponent}
Now we try to improve the 1PI NLO calculation of the dynamic critical exponent in model A \cite{Halperin:1972,Suzuki:1975}.
Though we perform the actual calculation based on the effective theory,
the calculation can be regarded as an approximation to the 2PI NLO calculation not only in the effective theory but also in the microscopic theory,
since the effective and microscopic theories are equivalent in the NLO 2PI $1/N$ expansion.
Our strategy is as follows:
Starting from the bare two-point function in the effective theory, $G^{E}_0(\p,p_0)^{-1} = \p^2 - i \Gamma_0^{-1}p_0$.
we first dress the two-point function only with the static self-energy
which is obtained in the 2PI NLO calculation to satisfy the static scaling
behavior at low momentum.
Then, using this dressed two-point function with the bare dynamic part,
$- i \Gamma_0^{-1}p_0$, as the modified ``bare'' propagator 
in the Schwinger-Dyson equation, we evaluate the infrared logarithmic term 
in $p_0$ and $\p$ in the dynamic part of the self-energy, 
from which we read off the dynamic critical exponent, $z$.
Thus, the difference of our approach from that of Ref.~ \cite{Halperin:1972,Suzuki:1975} lies in the inclusion of the static 2PI correlations
in the evaluation of the dynamic part.

Consider the Schwinger-Dyson equation,
\begin{align*}
  G(\p,p_0)^{-1} &= G_0(\p,p_0)^{-1} + \Sigma(\p,p_0).
\end{align*}
Separating the self-energy into the static and dynamic parts, $\Sigma(\p,p_0) = \Sigma(\p,0) + \Delta\Sigma(\p,p_0)$,
and combining the inverse of the bare response function and the static part of the self-energy, we rewrite the Schwinger-Dyson equation as
\begin{align*}
  G(\p,p_0)^{-1} &= \tilde G_0(\p,p_0) ^{-1}+ \Delta\Sigma(\p,p_0)
\end{align*}
where
\begin{align}
  \label{1.5PI}
  \tilde G_0(\p,p_0)^{-1} = G_0(\p,p_0)^{-1} + \Sigma(\p,0) = G(\p,0)^{-1} - i \Gamma_0^{-1}p_0.
\end{align}
Namely, the bare response function, $G_0$, is replaced by $\tilde G_0$, whose static part is that of the full response function but dynamic part is the bare response function.
This means $z=2-\eta$ at the tree level since $G^{-1}(\p,0) = \Lambda^\eta |\p|^{2-\eta}$.

Then we extract the $\log p_0$ term from the self-energy at zero momentum as in Ref.~\cite{Halperin:1972,Suzuki:1975}.
\begin{align}
  \label{self}
  \Delta \Sigma_{\rm NLO}(0,p_0) &= -\frac{4T}{N}p_0\int_{\q,q_0}{\rm Im}\tilde G_0(\q,p_0-q_0) \frac{1}{q_0(p_0-q_0)}
  \left\{ \tilde\Pi_0(\q,0)^{-1} - \tilde\Pi_0(\q,q_0)^{-1}\right\}.
\end{align}
By substituting Eq.~(\ref{1.5PI}) into Eq.~(\ref{self}) we obtain
\begin{align}
  \label{self'}
  \Delta \Sigma_{\rm NLO}(0,p_0) 
  &= -\frac{2T}{N}p_0\int_{\q} \frac{1}{p_0+i\gamma(\q)}G(\q,0) \left\{\tilde\Pi_0(\q,0)^{-1}-\tilde\Pi_0(\q,p_0+i\gamma(\q))^{-1}\right\},
\end{align}
where $G(\q,0) = \Lambda^{-\eta} |\q|^{\eta-2}$, $\gamma(\q) = \Gamma_0 G(\q,0)^{-1} = \Gamma_0 \Lambda^\eta |\q|^{2-\eta}$ and
\begin{align*}
  \tilde\Pi_0(\q,p_0)
  & = T\int_{{\boldsymbol k}} G({\boldsymbol k},0)G(\q-{\boldsymbol k},0)\frac{\gamma({\boldsymbol k})+\gamma(\q-{\boldsymbol k})}{\gamma({\boldsymbol k})+\gamma(\q-{\boldsymbol k})-ip_0}.
\end{align*}

In Eq.~(\ref{self'}), $\tilde\Pi_0(\q,0)$ and $\tilde\Pi_0(\q,p_0+i\gamma(\q))$ for low $p_0$ can be evaluated as
\begin{align}
   \label{tilde Pi_0}
   \tilde\Pi_0(\q,0) & = T\int_{{\boldsymbol k}} G({\boldsymbol k},0)G(\q-{\boldsymbol k},0) \notag \\
   & = T\Lambda^{-2\eta} |\q|^{2\eta-4+d} {\cal A}_d (\eta),
\end{align}
\begin{align}
   \label{tilde Pi_0'}
   \tilde\Pi_0(\q,p_0+i\gamma(\q)) & = T\int_{{\boldsymbol k}} G({\boldsymbol k},0)G(\q-{\boldsymbol k},0)\frac{\gamma({\boldsymbol k})+\gamma(\q-{\boldsymbol k})}{\gamma({\boldsymbol k})+\gamma(\q-{\boldsymbol k})+\gamma(\q)-ip_0} \notag \\
   & = T\Lambda^{-2\eta} |\q|^{2\eta-4+d} {\cal B}_d (\eta) \left(1 + {\cal O}\left(\frac{p_0}{\Gamma_0 \Lambda^\eta q^{2-\eta}}\right)\right),
\end{align}
where
\begin{align*}
  {\cal A}_d(\eta) = & \int_{\hat {\boldsymbol k}} |\hat {\boldsymbol k}|^{\eta-2}|\hat 1-\hat {\boldsymbol k}|^{\eta-2} 
  = \frac{1}{(4\pi)^{d/2}} \frac{\Gamma({2-\eta-\frac{1}{2}d})[\Gamma(\frac{1}{2}d+\frac{1}{2}\eta-1)]^2}{\Gamma(d+\eta-2)[\Gamma({1-\frac{1}{2}\eta})]^2},
\end{align*}
\begin{align*}
  {\cal B}_d(\eta) = \int_{\hat {\boldsymbol k}}
  \frac{|\hat {\boldsymbol k}|^{\eta-2} + |\hat 1-\hat {\boldsymbol k}|^{\eta-2}}
  {1 + |\hat {\boldsymbol k}|^{2-\eta} + |\hat 1-\hat {\boldsymbol k}|^{2-\eta}},
\end{align*}
and $\hat 1=\q/|\q|$ and $\hat {\boldsymbol k}={\boldsymbol k}/|\q|$.

Substituting Eqs.~(\ref{tilde Pi_0}) and (\ref{tilde Pi_0'}) into Eq.~(\ref{self'}) we obtain
\begin{align*}
  \Delta \Sigma_{\rm NLO}(0,p_0) = & -\frac{2T}{N}p_0\int_{\q} \frac{1}{p_0+i\gamma(\q)}G(\q,0) \left\{\tilde\Pi_0(\q,0)^{-1}-\tilde\Pi_0(\q,p_0+i\gamma(\q))^{-1}\right\} \notag \\
  = & -\frac{2T}{N}p_0\int_{\q} \frac{1}{p_0 + i \Gamma_0 \Lambda^\eta |\q|^{2-\eta}}\Lambda^{-\eta} |\q|^{\eta-2} \notag \\
  & \times \left\{\left(T\Lambda^{-2\eta} |\q|^{2\eta-4+d}{\cal A}_d(\eta)\right)^{-1} - \left(T\Lambda^{-2\eta} |\q|^{2\eta-4+d}{\cal B}_d(\eta)\left(1 + {\cal O}\left(\frac{p_0}{\Gamma_0 \Lambda^\eta |\q|^{2-\eta}}\right)\right)\right)^{-1}\right\} \notag \\
  = & i\frac{2}{N}\Gamma_0^{-1}p_0\int_{\q} \frac{|\q|^{2-\eta-d}}{|\q|^{2-\eta} - i\Gamma_0^{-1}\Lambda^{-\eta}p_0} \left({\cal A}_d(\eta)^{-1} - {\cal B}_d(\eta)^{-1}\left(1 + {\cal O}\left(\frac{p_0}{\Gamma_0 \Lambda^\eta |\q|^{2-\eta}}\right)\right)\right).
\end{align*}
From this we can extract the $\log p_0$ term in $\Delta \Sigma_{\rm NLO}(0,p_0)$ as 
\begin{align}
  \label{self_result}
  [\Delta \Sigma_{\rm NLO}(0,p_0)]_{\log p_0}
  = & i\frac{2}{N}\left({\cal A}_d(\eta)^{-1} - {\cal B}_d(\eta)^{-1}\right)\Gamma_0^{-1}p_0 \left[\frac{S_d}{(2\pi)^d} \int d|\q| \frac{|\q|^{1-\eta}}{|\q|^{2-\eta} - iC \Gamma_0^{-1} \Lambda^{-\eta} p_0}\right]_{\log p_0} \notag \\
  = & i\left({\cal B}_d(\eta)^{-1} - {\cal A}_d(\eta)^{-1}\right) \frac{S_d}{(2\pi)^d}\frac{1}{N}\Gamma_0^{-1}p_0\log p_0 \notag \\
  = & i\lambda\Gamma_0^{-1}p_0\log p_0,
\end{align}
where
\begin{align*}
  \lambda &= \left({\cal B}_d(\eta)^{-1} - {\cal A}_d(\eta)^{-1}\right)\frac{S_d}{(2\pi)^d}\frac{1}{N}
\end{align*}
and $S_d$ is the surface area of the $(d-1)$-dimensional unit sphere, $S_d=2\pi^{d/2}/\Gamma(d/2)$.
In Eq.~(\ref{self_result}), $C$ is a non-vanishing constant whose value is irrelevant for the calculation of the $\log p_0$ term.
Then, $z$ is obtained as
\begin{align}
\label{z_result}
   z = \frac{2-\eta}{1-\lambda} = \frac{2-\eta}{1- \left({\cal B}_d(\eta)^{-1} - {\cal A}_d(\eta)^{-1}\right)\frac{S_d}{(2\pi)^d}\frac{1}{N}},
\end{align}
where $\eta$ is the result of the NLO 2PI $1/N$ expansion, i.e.\ the solution of Eq.~(\ref{eta_2PI}).

Equation~(\ref{z_result}) is another main result of the present paper.
If one takes $\eta=0$ in the denominator, which amounts to using the free static response function, ${\cal A}_d(0)=1/8$ and ${\cal B}_d(0)=1/12$,
but if one substitutes the result of the NLO 1PI $1/N$ expansion, $\eta = \frac{8}{3\pi^2}\frac{1}{N}$ in the numerator, the result of Ref.~\cite{Halperin:1972,Suzuki:1975}, $z = 2 + \frac{4}{3\pi^2}\frac{1}{N}$, is reproduced.

Figure~2 shows the obtained result, $z$, as a function of $N$ for $d=3$.
The result of the NLO $1/N$ expansion in the effective theory, model A, \cite{Halperin:1972,Suzuki:1975} is also shown there.
One sees that our result approaches that of the strict $1/N$ expansion at NLO as $N$ increases but our result becomes smaller as $N$ decreases.
Also, our result has milder $N$-dependence than the previous result.
Our motivation, however, is academic, i.e.\ to theoretically explore possibility of improving the simple $1/N$ expansion, rather than to practically obtain more accurate results.
Therefore, we do not any more discuss how our result is better than the previous result.

\begin{center}
\begin{figure}[htbp]
  {\includegraphics[height=5cm]{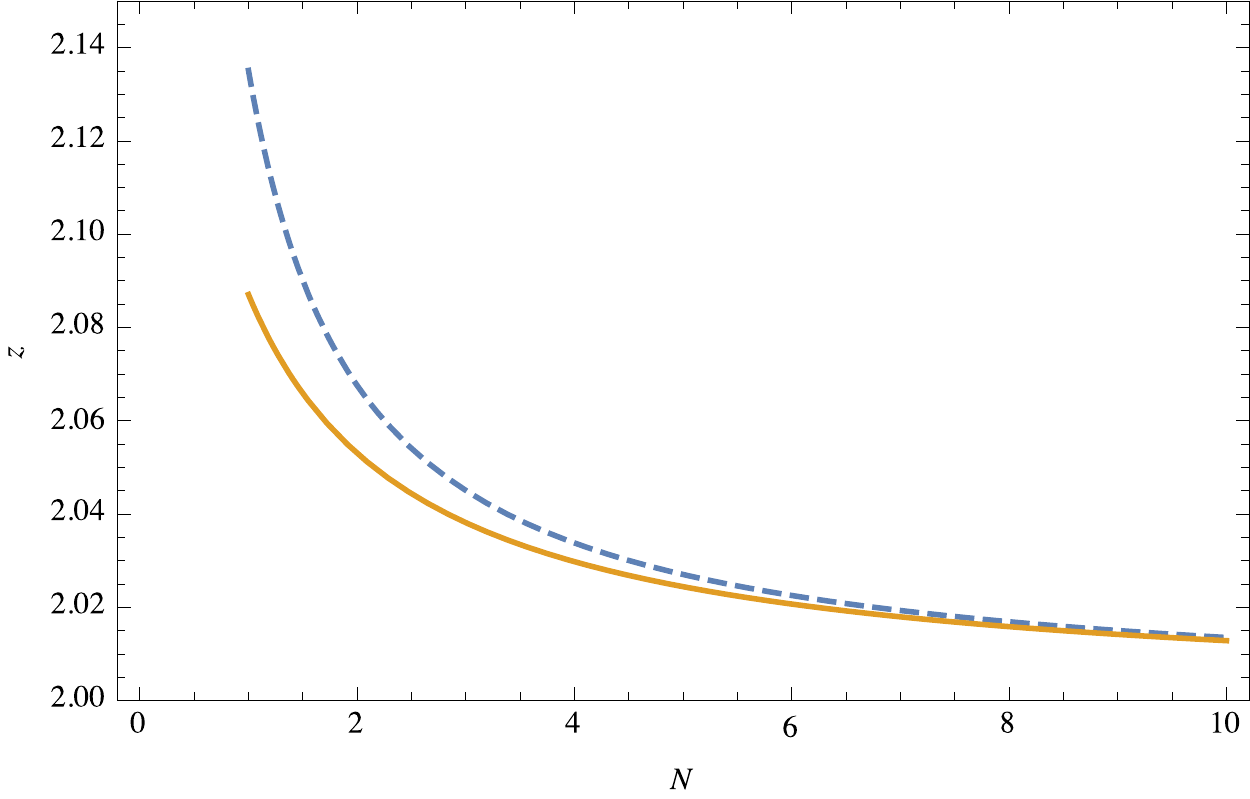}}
\caption{Dynamic critical exponent $z$ vs.\ $N$ for $d=3$. The solid curve is our result and the dashed curve is that of the NLO $1/N$ expansion in the effective theory, model A, 
\cite{Halperin:1972,Suzuki:1975}, $z = 2 + \frac{4}{3\pi^2}\frac{1}{N}$.}
\label{z}
\end{figure}
\end{center}

\section{summary and discussion}
In this paper we have studied the dynamic critical exponent from effective and microscopic theories.
We have employed a simple TDGL model, or model A in the classification of Ref.~\cite{Hohenberg:1977}, as an effective theory and
the imaginary time formalism of the finite-temperature filed theory as a microscopic theory.
Taking an $O(N)$ scalar model as an example and carrying out the $1/N$ expansion up to the NLO in the 1PI and 2PI effective actions,
we have compared the low-energy and low-momentum behavior of the response function in the effective theory and of the retarded Green's function in the microscopic theory.

The results are summarized in Table 1.
On the one hand, at the NLO of the 1PI $1/N$ expansion the low-energy and low-momentum behavior of the two-point function is  very much different in the microscopic and effective theories:
in the field theory it is dominated by the propagating mode while in model A it is dominated by the diffusive mode.
 Also, in the microscopic theory the dynamic critical exponent, $z$, depends on whether the kinematics is relativistic or nonrelativistic.
On the other hand, at the NLO of the 2PI $1/N$ expansion the microscopic and effective theories are equivalent.
They satisfy exactly the same Kadanoff-Baym equation.
Also, whether the kinematics is relativistic or nonrelativistic in the microscopic theory becomes irrelevant.
This implies that the diffusive mode with $z = 2 + {\cal O}(1/N)$ is dominant at low energies and momenta even in the microscopic theory at the NLO of the 2PI $1/N$ expansion,
though we have not explicitly solved the Kadanoff-Baym equation.

\begin{center}
\begin{table}[h]
\caption{Summary of static and dynamic critical exponents for $d=3$ at the NLO of 1PI and 2PI $1/N$ expansion in the effective and microscopic theories.}
\begin{tabular}{|c|c|c|c|c|c|c|}
\hline
  \multirow{3}{*}{\backslashbox{exponent}{theory}} &  \multicolumn{3}{c|}{1PI} & \multicolumn{3}{c|} {2PI} \\
  \cline{2-7}
  &  \multirow{2}{*}{effective theory} & \multicolumn{2}{c|}{microscopic theory} & \multirow{2}{*}{effective theory} & \multicolumn{2}{c|}{microscopic theory} \\
  \cline{3-4}
  \cline{6-7}
  & & nonrelativistic & relativistic & & nonrelativistic & relativistic \\
\hline
  static critical exponent & \multicolumn{3}{c|}{Ref.~\cite{Ma:1973}} & \multicolumn{3}{c|}{Ref.~\cite{Alford:2004jj}} \\
  $\eta$ & \multicolumn{3}{c|}{$\frac{8}{3\pi^2}\frac{1}{N}$} & \multicolumn{3}{c|}{Fig.~1} \\
  & \multicolumn{3}{c|}{} & \multicolumn{3}{c|}{} \\
\hline
  dynamic critical exponent & Ref.~\cite{Halperin:1972,Suzuki:1975} & Ref.~\cite{Kondor:1974,Abe:1973,Abe:1974,Suzuki:1974} & present paper & \multicolumn{3}{c|}{present paper} \\
  $z$ & $2+\frac{4}{3\pi^2}\frac{1}{N}$ & $2-\frac{32}{3\pi^2}\frac{1}{N}$ & $1-\frac{16}{3\pi^2}\frac{1}{N}$ & \multicolumn{3}{c|}{Fig.~2 (approximate solution)} \\
  & (diffusive) & (propagating) & (propagating) & \multicolumn{3}{c|}{(diffusive)} \\
\hline
\end{tabular}
\end{table}
\end{center}

We have also tried to improve the calculation of the dynamic critical exponent of model A by incorporating the static 2PI NLO correlations.
This calculation can be regarded as an approximation to the 2PI NLO calculation of the dynamic critical exponent not only in the effective theory but also in the microscopic theory.
By incorporating the static 2PI correlations into the two-point function we identify the logarithmic term at low energies and momenta in the $1/N$ NLO self-energy, 
 from which we determine the critical exponent, $z$.
The obtained critical exponent is slightly smaller than the previous result and its $N$ dependence is also milder than the previous one. 

To explicitly solve the Kadanoff-Baym equation is our future problem.

One comment is in order here.
In the microscopic $O(N)$ scalar theory the energy and $O(N)$ charges are conserved.
If these conserved quantities couple to the order parameter, the corresponding model in the classification of Ref.~\cite{Hohenberg:1977} is different from model A.
The effect of the coupling to the $O(N)$ charges does not show up at the NLO $1/N$ expansion so that we have to go to higher orders.
The effect of the coupling to the conserved energy also requires further studies.
These are our future problems as well.

\section*{Acknowledgments}
The authors would like to thank J\"urgen Berges 
for helpful discussions and correspondence 
on critical dynamics of $O(N)$ scalar theory. 

\appendix
\section{Expressions for general bubble diagrams}
Consider a general bubble diagram of $A$ and $B$ in the effective and microscopic theories,
where $A$ and $B$ can be either elementary or composite.

In the TDGL theory
\begin{align*}
\label{}
  \Pi^{E}_{AB}(\p,p_0)
  &= \int_{\q,q_0} \left\{ G^{E}_A(\p-\q,p_0-q_0)C^{E}_B(\q,q_0)
  + C^{E}_A(\p-\q,p_0-q_0)G^{E}_B(\q,q_0) \right\} \\
  &= \int_{\q,q_0} \left\{G^{E}_A(\p-\q,p_0-q_0) \frac{2T}{q_0}{\rm Im}G^{E}_B(\q,q_0)
  + \frac{2T}{p_0-q_0}{\rm Im}G^{E}_A(\p-\q,p_0-q_0) G^{E}_B(\q,q_0)\right\}.
\end{align*}

In the imaginary-time formalism of the field theory at finite temperature
\begin{align*}
\label{}
  \Pi^{M}_{AB}(\p,i\omega_n) &= T\sum_m\int_\q G^{M}_A(\p-\q,i\omega_n-i\omega_m) G^{M}_B(\q,i\omega_m) \\
  &= T\sum_m\int_\q \int \frac{d{{E}}'}{\pi} \frac{{\rm Im}G^{M}_A(\p-\q,{{E}}')}{{{E}}'-i\omega_n+i\omega_m} \int \frac{d{{E}}}{\pi} \frac{{\rm Im}G^{M}_B(\q,{{E}})}{{{E}}-i\omega_m} \\
  &= T\sum_m\int_\q \int \frac{d{{E}}'}{\pi} \frac{d{{E}}}{\pi} {\rm Im}G^{M}_A(\p-\q,{{E}}'){\rm Im}G^{M}_B(\q,{{E}}) 
  \left(\frac{1}{{{E}}'-i\omega_n+i\omega_m}+\frac{1}{{{E}}-i\omega_m}\right)\frac{1}{{{E}}+{{E}}'- i\omega_n} \\
  &= \int_\q \int \frac{d{{E}}'}{\pi} \frac{d{{E}}}{\pi} {\rm Im}G^{M}_A(\p-\q,{{E}}'){\rm Im}G^{M}_B(\q,{{E}}) 
  \left(-f^M(-{{E}}')+f^M({{E}})\right)\frac{1}{{{E}}+ {{E}}'- i\omega_n} \\
  &= \int_\q\int \frac{d{{E}}}{\pi} G^{M}_A(\p-\q,i\omega_n-{{E}})f^M({{E}}){\rm Im}G^{M}_B(\q,{{E}}) \\
  &\phantom{=} - \int_\q\int \frac{d{{E}}'}{\pi} f^M(-{{E}}'){\rm Im}G^{M}_A(\p-\q,{{E}}')G^{M}_B(\q,i\omega_n - {{E}}'),
\end{align*}
where $f^M({{E}})=[\exp({{E}}/T)-1]^{-1}$.
In this equation $G^{M}_A$ and $G^{M}_B$ are analytically continued to the complex energy plane and
should be understood as retarded Green's functions in ${\rm Im}G^{M}_A$ and ${\rm Im}G^{M}_B$. 
Letting $i\omega_n \rightarrow p_0+ i\epsilon$, we obtain
\begin{align*}
\label{}
  \Pi^{M}_{AB}(\p,p_0)
  &= \int_\q\int \frac{d{{E}}}{\pi} G^{M}_A(\p-\q,p_0 - {{E}})f^M({{E}}){\rm Im}G^{M}_B(\q,{{E}}) \\
  &\phantom{=} - \int_\q\int \frac{d{{E}}'}{\pi} f^M(-{{E}}'){\rm Im}G^{M}_A(\p-\q,{{E}}')G^{M}_B(\q,p_0 - {{E}}') \\
  &= \int_{\q,q_0} \left\{ G^{M}_A(\p-\q,p_0-q_0)2f^M(q_0){\rm Im}G^{M}_B(\q,q_0) \right. \\
  &\phantom{= \int_{\q,q_0}} \left.- 2f^M(q_0 - p_0){\rm Im}G^{M}_A(\p-\q,p_0-q_0)G^{M}_B(\q,q_0) \right\}.
\end{align*}
Again, the Green's functions should be understood as the retarded ones in this and later equations.

\section{Separation of the static and dynamic parts of the bubble diagram}
Consider a general bubble diagram $\Pi_{AB}(\p,p_0)$ of $A$ and $B$.
\begin{align}
\label{bubble}
  \Pi_{AB}(\p,p_0) &= \int_{\q,q_0}
  \left\{ \frac{2T}{p_0-q_0}{\rm Im}G_A(\p-\q,p_0-q_0) G_B(\q,q_0)
  + G_A(\p-\q,p_0-q_0) \frac{2T}{q_0}{\rm Im}G_B(\q,q_0)\right\}.
\end{align}
We show that $\Pi_{AB}(\p,p_0)$ can be decomposed into the static part $\Pi_{AB}(\p,0)$ and the dynamic part $\Delta \Pi_{AB}(\p,p_0)$ as follows:
\begin{align*}
\label{}
  \Pi_{AB}(\p,p_0) &= \Pi_{AB}(\p,0) + \Delta \Pi_{AB}(\p,p_0) \\
  &= T \int_{\q} G_A(\p-\q,0)G_B(\q,0) \\
  &-2 T p_0 \int_{\q,q_0} {\rm Im}G_A(\p-\q,p_0-q_0)
  \frac{1}{q_0(p_0-q_0)}\left\{G_B(\q,0)-G_B(\q,q_0)\right\}.
\end{align*}
By repeatedly using the spectral representation of $G_A(\p,p_0)$ and $G_B(\p,p_0)$:
\begin{align*}
\label{}
  G_A(\p,p_0) = & \frac{1}{\pi}\int dq_0 \frac{{\rm Im}G_A(\p,q_0)}{q_0-p_0-i\epsilon}, \\
  G_B(\p,p_0) = & \frac{1}{\pi}\int dq_0 \frac{{\rm Im}G_B(\p,q_0)}{q_0-p_0-i\epsilon},
\end{align*}
we can rewrite the second term of Eq.~(\ref{bubble}) as follows,
\begin{align*}
\label{}
  &\int_{\q,q_0} G_A(\p-\q,p_0-q_0) \frac{2T}{q_0}{\rm Im}G_B(\q,q_0) \\
  =&2T \int_{\q,q_0} \frac{1}{\pi}\int dq_0' \frac{{\rm Im}G_A(\p-\q,q_0')}{q_0'-p_0+q_0-i\epsilon}
  \frac{1}{q_0}{\rm Im}G_B(\q,q_0) \\
  =&2T \int_{\q,q_0} \frac{1}{\pi}\int dq_0'
  {\rm Im}G_A(\p-\q,q_0')\frac{1}{q_0'-p_0}
  \left(\frac{1}{q_0}-\frac{1}{q_0'-p_0+q_0-i\epsilon}\right) {\rm Im}G_B(\q,q_0) \\
  =&2T \int_{\q,q_0'} {\rm Im}G_A(\p-\q,q_0')\frac{1}{q_0'-p_0}
  \left\{G_B(\q,0)-G_B(\q,p_0-q_0')\right\} \\
  =&2T \int_{\q,q_0'} {\rm Im}G_A(\p-\q,q_0')\left\{\frac{1}{q_0'}G_B(\q,0)
  +\left(\frac{1}{q_0'-p_0}-\frac{1}{q_0'}\right)G_B(\q,0)
  -\frac{1}{q_0'-p_0}G_B(\q,p_0-q_0')\right\} \\
  =& T \int_\q G_A(\p-\q,0)G_B(\q,0) \\
  +&2T \int_{\q,q_0} {\rm Im}G_A(\p-\q,p_0-q_0)\left\{
  -\left(\frac{1}{q_0}+\frac{1}{p_0-q_0}\right)G_B(\q,0)
  +\frac{1}{q_0}G_B(\q,q_0)\right\}. \\
\end{align*}
By combining with the first term we obtain the relation
\begin{align*}
\label{}
  \Pi_{AB}(\p,p_0) &= T \int_\q G_A(\p-\q,0)G_B(\q,0) \\
  &-2 T p_0 \int_{\q,q_0} {\rm Im}G_A(\p-\q,p_0-q_0)
  \frac{1}{q_0(p_0-q_0)}\left\{G_B(\q,0)-G_B(\q,q_0)\right\}.
\end{align*}

\section{Calculation of $\Delta\Sigma^{M}_{\rm NLO}$ at the NLO of the 1PI $1/N$ expansion}
Consider the dynamic part of the self-energy at the NLO of the 1PI 1$/N$ expansion in the microscopic theory, Ref.~\cite{Boyanovsky:2000nt,Boyanovsky:2001pa},
\begin{align}
\label{C1}
  \Delta \Sigma^{M}_{\rm NLO}(\p,p_0) = \frac{2u T}{3N}p_0\int_{\q,q_0}
  {\rm Im}G^{M}_0(\p-\q,p_0-q_0) \frac{1}{q_0(p_0-q_0)} \left\{ D^{M}_0(\q,0) - D^{M}_0(\q,q_0)\right\},
\end{align}
where $D^{M}_0(\q,q_0)$ is given by Eq.~(\ref{D_0}).

If $|q_0|, |\q| \ll T$,
\begin{align}
  \label{C2}
  \Pi^{M}_0 (\q,q_0) = \frac{T}{8|\q|}\left\{\theta(|\q|^2-q_0^2)+i\frac{1}{\pi}\log\left|q_0 + |\q| \over q_0 - |\q|\right|\right\}.
\end{align}
Substituting
\begin{align*}
  {\rm Im}G^{M}_0(\p,p_0) &= \frac{\pi}{2|\p|}\left(\delta(p_0-|\p|)-\delta(p_0+|\p|)\right)
\end{align*}
into Eq.~(\ref{C1}) leads to
\begin{align}
\label{C3}
  \Delta \Sigma^{M}_{\rm NLO}(\p,p_0) 
  &= \frac{4u T}{3N}p_0^2\int_\q \frac{1}{4|\p-\q|^2}
  \frac{1}{p_0^2-|\p-\q|^2} D^{M}_0(\q,0) \nonumber\\
  &-\frac{2u T}{3N}p_0\int_\q \frac{1}{4|\p-\q|^2}\left\{\frac{1}{p_0-|\p-\q|}D^{M}_0(\q,p_0-|\p-\q|)
  +\frac{1}{p_0+|\p-\q|}D^{M}_0(\q,p_0+|\p-\q|)\right\}.
\end{align}

We first examine the logarithmic contribution in $\Delta \Sigma^{M}_{\rm NLO}$.
Consider the term including $D^{M}_0(\q,0)$ in Eq.~(\ref{C3}).
If $|p_0|, |\p| \ll T$, the integral is dominated by low $|\q|$.
At low $|\q|$, $\Pi^{M}_0(\q,0)$ diverges as $1/|\q|$ and one can neglect $1$ in the denominator of $D^{M}_0(\q,0)$ in Eq.~(\ref{D_0}).
Then,
\begin{align*}
\label{}
  \frac{4u T}{3N}p_0^2\int_\q \frac{1}{4|\p-\q|^2}\frac{1}{p_0^2-|\p-\q|^2} D^{M}_0(\q,0)
  &= \frac{8T}{N}p_0^2\int_\q \frac{1}{4|\p-\q|^2}\frac{1}{p_0^2-|\p-\q|^2} \Pi^{M}_0(\q,0)^{-1} \\
  &= \frac{8T}{N}p_0^2\int_\q \frac{1}{4|\p-\q|^2}\frac{1}{p_0^2-|\p-\q|^2} \frac{8|\q|}{T},
\intertext{which becomes, if $|\p| \ll |p_0|$,}
  &= \frac{8T}{N}p_0^2\int_\q \frac{1}{4\q^2}\frac{1}{p_0^2-\q^2} \frac{8|\q|}{T} \\
  &= \frac{8}{\pi^2}\frac{1}{N} p_0^2\log p_0,
\intertext{and if $|p_0| \ll |\p|$,}
  &= -\frac{8T}{N}p_0^2\int_\q \frac{1}{4|\p-\q|^4} \frac{8|\q|}{T} \\
  &= \frac{8}{\pi^2}\frac{1}{N} p_0^2\log |\p|.
\end{align*}
The terms including $D^{M}_0(\q,p_0-|\p-\q|)$ or $D^{M}_0(\q,p_0+|\p-\q|)$ generate no logarithmic contribution:
In $D^{M}_0(\q,p_0-|\p-\q|)$ or $D^{M}_0(\q,p_0+|\p-\q|)$, $1/|\q|$ behavior in $D^{M}_0(\q,0)$ at low $|\q|$ is cut off as can be seen in Eq.~(\ref{C3}).
Therefore, the logarithmic term of $\Delta \Sigma^{M}_{\rm NLO}$ is given as
\begin{align*}
\label{}
  \Delta \Sigma^{M}_{\rm NLO}(\p,p_0) = 
  \begin{cases}
  \displaystyle{\frac{8}{\pi^2}\frac{1}{N} p_0^2\log p_0} & (|\p| \ll |p_0|) \\
  \displaystyle{\frac{8}{\pi^2}\frac{1}{N} p_0^2\log |\p|} & (|p_0| \ll |\p|)
  \end{cases},
\end{align*}
which is real.

Having observed no logarithmic contribution in the imaginary part of $\Delta \Sigma^{M}_{\rm NLO}$,
we determine the dominant contribution of {\rm Im}$\Delta \Sigma^{M}_{\rm NLO}$ at low $|p_0|$ and $|\p|$.
The term including $D^{M}_0(\q,0)$ does not contribute and
\begin{align*}
\label{}
  {\rm Im}\Delta \Sigma^{M}_{\rm NLO}(\p,p_0) 
  =-\frac{2u T}{3N}p_0 \int_\q \frac{1}{4|\p-\q|^2}&\left\{\frac{1}{p_0-|\p-\q|}{\rm Im}D^{M}_0(\q,p_0-|\p-\q|)\right\} \\
  &+\left.\frac{1}{p_0+|\p-\q|}{\rm Im}D^{M}_0(\q,p_0+|\p-\q|)\right\}.
\end{align*}
The imaginary part of $\Pi^{M}_0(\q,q_0)$ is logarithmically divergent when one approaches the light-cone, $q_0 \pm |\q| \rightarrow 0$.
Therefore, in order to determine  the dominant contribution of {\rm Im}$\Delta \Sigma^{M}_{\rm NLO}$ at low $|p_0|$ and $|\p|$,
one can replace as follows,
\begin{align*}
  {\rm Im}\Pi^{M}_0(\q,p_0\mp|\p-\q|) &= \frac{T}{8\pi |\q|}\log\left|p_0\mp|\p-\q| + |\q| \over p_0\mp|\p-\q| - |\q|\right| \\
  &=
  \begin{cases}
    \displaystyle{\pm\frac{T}{8\pi |\q|}\log|p_0|} & (|\p| \ll |p_0|) \\
    \displaystyle{\pm\frac{T}{8\pi |\q|}\log|\p|} & (|p_0| \ll |\p|)
  \end{cases},
  \intertext{and}  
  {\rm Im}D^{M}_0(\q,p_0\mp|\p-\q|)&=
  \begin{cases}
  \displaystyle{\pm\frac{6}{u}\frac{8\pi |\q|}{T}\frac{1}{\log|p_0|}} & (|\p| \ll |p_0|) \\
  \displaystyle{\pm\frac{6}{u}\frac{8\pi |\q|}{T}\frac{1}{\log|\p|}} & (|p_0| \ll |\p|)
\end{cases}.
\end{align*}
Substituting the above into Eq.~(\ref{C3}), we obtain if $|\p| \ll |p_0|$,
\begin{align*}
\label{}
  {\rm Im}\Delta \Sigma^{M}_{\rm NLO}(\p,p_0) 
  &=-\frac{2u T}{3N}p_0\int_\q \frac{1}{4\q^2}\left\{\frac{1}{p_0-|\q|}{\rm Im}D^{M}_0(\q,p_0-|\q|)
  +\frac{1}{p_0+|\q|}{\rm Im}D^{M}_0(\q,p_0+|\q|)\right\} \\
  &=-\frac{16\pi}{N}p_0\frac{1}{\log|p_0|}\int_\q \frac{1}{p_0^2-\q^2} \\
  &=-\frac{16\pi}{N}p_0|p_0|\frac{1}{\log|p_0|}\int_{\hat\q} \frac{1}{1-\hat\q^2},
\end{align*}
and if $|p_0| \ll |\p|$,
\begin{align*}
\label{}
  {\rm Im}\Delta \Sigma^{M}_{\rm NLO}(\p,p_0) 
  &=-\frac{2u T}{3N}p_0\int_\q \frac{1}{4|\p-\q|^3}\left\{-{\rm Im}D^{M}_0(\q,-|\p-\q|)
  +{\rm Im}D^{M}_0(\q,|\p-\q|)\right\} \\
  &=-\frac{16\pi}{N}p_0\frac{1}{\log|\p|}\int_{\q} \frac{|\q|}{|\p-\q|^3} \\
  &=-\frac{16\pi}{N}p_0|\p|\frac{1}{\log|\p|}\int_{\hat\q} \frac{|\hat \q|}{|\hat 1-\hat \q|^3},
\end{align*}
where $\hat 1=\p/|\p|$ and $\hat \q=\q/|\p|$.

\end{document}